\documentclass[12pt,letterpaper,english,aps]{revtex4}
\usepackage{times}
\usepackage[T1]{fontenc}
\usepackage[latin1]{inputenc}
\usepackage{subfigure}
\usepackage{graphicx}
\usepackage{amssymb}

\makeatletter

\providecommand{\tabularnewline}{\\}

\makeatother
\begin{document}

\title{The convolution theorem for nonlinear optics}

\author{Hernando Garcia%
\thanks{contact author at hgarcia@siue.edu%
}}

\email{hgarcia@siue.edu}

\affiliation{Department of Physics Southern Illinois University Edwardsville,
IL 62026 }

\author{Ramki Kalyanaraman }

\affiliation{Department of Physics Washington University in St. Louis St. Louis,
MO 63130 }

\begin{abstract}
We have expressed the nonlinear optical absorption of a semiconductor
in terms of its linear spectrum. We determined that the two-photon
absorption coefficient in a strong DC-electric field of a direct gap
semiconductor can be expressed as the product of a differential operator
times the convolution integral of the linear absorption without a
DC-electric field and an Airy function. We have applied this formalism
to calculate the two-photon absorption coefficient and nonlinear refraction
for GaAs and ZnSe using their linear absorption and have found excellent
agreement with available experimental data. 
\end{abstract}
\maketitle
A fundamental limitation in non-linear spectroscopy is the requirement
for large peak laser intensities because a coherent $N$-photon process
($N\geq2$) has a small cross section. With the limited availability
of continuous laser sources having broad bandwidths and good coherence,
nonlinear spectroscopy is challenging. In contrast linear absorption
cross sections are much larger, especially in semiconductors close
to a critical point (Van Hove singularities). Therefore, it would
be ideal if the nonlinear properties, such as the two-photon absorption
and nonlinear refraction of a semiconductor, could be predicted from
their linear spectrum, which is relatively straightforward to obtain.
In this Letter, we present a theoretical approach to predict the two-photon
absorption spectrum of a direct gap semiconductor based only on its
linear absorption spectrum close to the band-edge. The formalism developed
here also gives information about the role of a DC-electric field
on the nonlinear optical response of semiconductors. We have also
applied the Kramers-Kronig relation to calculate the nonlinear refraction
and have obtained excellent agreement with the available experimental
data for $GaAs$ and $ZnSe$ direct-gap semiconductors. This theory
could be of great significance towards identifying promising nonlinear
optical materials for application in diverse areas such as optical
switching and optical limiting. 

The effect of electric field on the dielectric constant of solids
has been extensively investigated in the past \cite{tharmalingam63,morgan66,yacoby68,aspnes67,enderlein67,french68}.
The effect, known as the Franz-Keldysh (FK) effect \cite{franz58,keldysh65},
has been used as a tool in spectroscopy to modulate the energy gap
and resolve details of the band structure otherwise embedded in a
broadband background \cite{aspnes70,aspnes80,cardona96,seraphin65}.
Recently, we reported the calculation of the nonlinear absorption
coefficient in the presence of a very strong electric field for direct
as well as indirect gap semiconductors and extended the formalism
to the N-photon process \cite{garcia06d,garcia06a}. At the heart
of the calculation is the use of a modified Volkov wavefunction that
includes the effect of the electric field in one direction (Airy Function),
and uses the S-matrix to calculate the N-photon transition rate in
first-order perturbation theory. We worked in the effective mass approximation
and assumed that the momentum matrix elements were independent of
electron-hole wave vector k. We also assumed that the optical field
only modified the final energy of the electron-hole pair. Finally,
we considered an isotropic solid with a full valence band and an empty
conduction band. The resulting generalized N-photon absorption coefficient
in the presence of a DC-electric field was given by \cite{garcia06d}:
\begin{equation}
\beta^{(N)}=\frac{\alpha_{b}f^{1/3}}{2^{2N-1}\pi}\left[\frac{8\pi e^{2}E_{\mu}}{n_{o}m_{c}^{2}c\omega^{4}}\right]^{N-1}\left[\frac{N(2N-3)!!}{((N-1)!)^{3}}\right]\left(\frac{m_{\mu}}{\hbar^{2}}\right)^{N}\int_{\epsilon_{o}^{(N)}}^{\infty}(\epsilon-\epsilon_{o}^{(N)})^{N-1}\mid Ai(\epsilon)^{2}\mid d\epsilon\label{eq:betaN}\end{equation}
where $E_{\mu}=\left(\hbar^{2}e^{2}F^{2}/2m_{\mu}\right)^{1/3}$ is
the characteristic energy of the DC electric field $F$, $m_{\mu}$
is the electron effective mass, $m_{c}$ is mass of the electron in
the conduction band, $n_{o}$ is the semiconductor index of refraction,
$Ai(\epsilon)$ is the Airy function, and $f$, $\alpha_{b}$, and
$\epsilon_{o}^{(N)}$ are given by: $f=\frac{2eFm_{\mu}}{\hbar^{2}}$;
$\alpha_{b}=\frac{8\pi^{2}}{n_{o}c}\frac{\mid P_{vc}\mid^{2}e^{2}}{m_{o}^{2}\omega}$;
and $\epsilon_{o}^{(N)}=\frac{E_{g}-N\hbar\omega}{E_{\mu}}$, where
$m_{o}$ is the electron bare mass, and $P_{vc}$ is the interband
momentum matrix elements. Using the following property of the Airy
function \cite{aspnes67}:\begin{equation}
\int_{o}^{\infty}t^{n}\mid Ai(t+x)\mid^{2}dt=\frac{n}{2n+1}\left[\frac{1}{2}\frac{d^{2}}{dx^{2}}-2x\right]\int_{o}^{\infty}t^{n-1}\mid Ai(t+x)\mid^{2}dt\label{eq:Ai}\end{equation}
the integral in Eq. \ref{eq:betaN} can be reduced after successive
applications of Eq. \ref{eq:Ai} and using the below property:\begin{equation}
\int_{0}^{\infty}\mid Ai(t+\epsilon_{o}^{(N)}\mid^{2}dt=\frac{2.4^{2/3}}{\pi}\int_{0}^{\infty}t^{\frac{1}{2}}Ai\left(t+\frac{E_{g}-N\hbar\omega}{\Omega}\right)dt\label{eq:Aieq2}\end{equation}
obtained from \cite{abramowitz}, where $\Omega=4^{-1/3}E_{\mu}$.
As a consequence, the $N$-photon absorption coefficient $\beta^{(N)}$,
Eq. \ref{eq:betaN}, can be expressed in terms of the linear absorption
as:

\begin{eqnarray}
\beta^{(N)} & = & \frac{1}{2^{2N-1}\omega}\left[\frac{8\pi e^{2}E_{\mu}m_{\mu}}{\hbar^{2}n_{o}m_{c}^{2}c\omega^{4}}\right]^{N-1}\left[\frac{N(2N-3)!!}{\left[(N-1)!\right]^{2}\prod_{i=1}^{N}(2i-1)}\right]\left[\frac{1}{2}\frac{d^{2}}{d\epsilon_{o}^{(N)2}}-2\epsilon_{o}^{(N)}\right]^{N-1}\times\nonumber \\
 &  & \int\beta^{(1)}(E^{'})\frac{\omega^{'}}{\Omega}Ai\left(\frac{E^{'}-N\hbar\omega}{\Omega}\right)dE^{'}\label{eq:betaN2}\end{eqnarray}
where $\beta^{(1)}$ is given by: \begin{equation}
\beta^{(1)}=\frac{\alpha_{b}}{(2\pi)^{2}}\left[\frac{2m_{\mu}}{\hbar^{2}}\right]^{3/2}(E^{'}-E_{g})^{1/2}=\frac{\alpha_{b}^{'}(2m)^{3/2}}{(2\pi)^{2}\hbar^{2}}\frac{1}{E^{'}}(E^{'}-E_{g})^{1/2}\label{eq:beta1}\end{equation}
where we have redefined $\alpha_{b}$ such that the new $\alpha_{b}^{'}$
and $E_{g}$ can be used as fitting parameters for the linear absorption.
We see that Eq. \ref{eq:betaN2} reduces to the well-know convolution
theorem for $N=1$ \cite{aspnes68}. \emph{We call Eq.} \ref{eq:betaN2}
\emph{the $N$-photon absorption convolution theorem (i.e. the nonlinear
convolution theorem) and view the differential operator as the $N$-photon
absorption operator}. In the case of $N=2$, the two-photon absorption
coefficient is given by:\begin{equation}
\beta^{(2)}=\frac{2}{3}\frac{\pi e^{2}E_{\mu}m_{\mu}}{\hbar^{2}n_{o}m_{c}^{2}c\omega^{5}}\left[\frac{1}{2}\frac{d^{2}}{d\epsilon_{o}^{(2)2}}-2\epsilon_{o}^{(2)}\right]\int_{-\infty}^{\infty}\beta^{(1)}(E^{'})\frac{\omega^{'}}{\Omega}Ai\left(\frac{E^{'}-2\hbar\omega}{\Omega}\right)dE^{'}\label{eq:beta2}\end{equation}
Eq. \ref{eq:beta2} contains a very remarkable result: \emph{the two-photon
absorption is given by a convolution of the linear absorption}. So
if the spectrum of the linear absorption close to the band edge is
known/measured then Eq. \ref{eq:beta2} can be used to generate the
nonlinear absorption spectrum of the semiconductor. Also, using the
familiar Kramers-Kronig (\emph{KK}) relationship for nonlinear optics
\cite{sheikbahae91} along with Eq. \ref{eq:beta2} we get: \begin{equation}
n^{(2)}(\omega,F)=\frac{c}{\pi}\int_{0}^{\infty}\frac{\beta^{(2)}(\omega,F)}{\omega^{'2}-\omega^{2}}d\omega^{'}\label{eq:n2}\end{equation}
for the nonlinear refraction. 

Recently, there have been reports on the development of a new technique
to measure the nonlinear absorption in a broad spectral range using
Z-scan and a supercontinuum laser source \cite{balu05,he02}. Using
this technique, the spectral distribution of the two-photon absorption
for $ZnSe$ was measured. To test the above theory, we have calculated
the two-photon absorption for $GaAs$ and $ZnSe$ semiconductors using
Eq. \ref{eq:beta2}. First, as shown in Fig. \ref{fig:GaAs-data}(a),
the absorption coefficient for $GaAs$ was estimated using Eq. \ref{eq:beta1}
and using $\alpha_{b}^{'}$ and $E_{g}$ as fitting parameters, with
low temperature experimental values taken from \cite{palik85} (where
the absorption edge is dominant). Table. \ref{tab:Table-of-quantities}
shows the values used in our calculation and the results for the energy
gap and $\alpha_{b}^{'}$. From the above result \emph{}we calculated
the two-photon absorption and the nonlinear refraction and compared
it to the experimentally measured values \cite{stryland85,hurlbut07},
as shown in Fig. \ref{fig:GaAs-data}(b) and Fig. \ref{fig:GaAs-data}(c).
We have done a similar analysis for $ZnSe$ and the results for the
linear absorption are shown in Fig. \ref{fig:ZnSe-data}(a), with
experimental data taken from \cite{SopraDataBase}. The two-photon
absorption and nonlinear refraction are shown in Fig. \ref{fig:ZnSe-data}(b)
and (c) respectively, with experimental values taken from \cite{balu05}.
In our calculations, we have used a characteristic energy of the DC
field $E_{\mu}=1/10\times E_{g}$ in order to minimize band bending
and allow comparison with the zero field case ($F=0$). The excellent
agreement is indeed quite remarkable, especially when considering
that the experimental absorption data for $ZnSe$ was taken from known
values of $n$ and $k$ at room temperature \cite{SopraDataBase},
and related to the absorption coefficient through $\alpha=(\frac{\omega}{n_{O}c})Im(\epsilon),$
where $\epsilon=n^{2}+k^{2}+2ink$, is the complex dielectric function.

In conclusion, we have explored some of the mathematical structures
of the N-photon absorption process in the presence of a very strong
DC-field. We have found that the nonlinear absorption can be expressed
as the product of an N-photon operator times the linear absorption
coefficient. This is, to our knowledge, the first time that nonlinear
processes have been viewed as a consequence of a single photon process
rescaled to an energy gap given by $E_{g}/N$. We also found that
because of this relation, results such as the convolution theorem,
and \emph{KK} can be introduced naturally. Finally, we applied this
formalism to two well known semiconductors ($GaAs$ and $ZnSe$) and
found excellent agreement with experimental measured trends. This
approach can be of great value in predicting nonlinear properties
solely from measurements of linear properties.

RK acknowledges support by the National Science Foundation through
grant \# DMI-0449258.


\begin{thebibliography}{23}
\expandafter\ifx\csname natexlab\endcsname\relax\def\natexlab#1{#1}\fi
\expandafter\ifx\csname bibnamefont\endcsname\relax
  \def\bibnamefont#1{#1}\fi
\expandafter\ifx\csname bibfnamefont\endcsname\relax
  \def\bibfnamefont#1{#1}\fi
\expandafter\ifx\csname citenamefont\endcsname\relax
  \def\citenamefont#1{#1}\fi
\expandafter\ifx\csname url\endcsname\relax
  \def\url#1{\texttt{#1}}\fi
\expandafter\ifx\csname urlprefix\endcsname\relax\def\urlprefix{URL }\fi
\providecommand{\bibinfo}[2]{#2}
\providecommand{\eprint}[2][]{\url{#2}}

\bibitem[{\citenamefont{Tharmalingam}(1963)}]{tharmalingam63}
\bibinfo{author}{\bibfnamefont{K.}~\bibnamefont{Tharmalingam}},
  \bibinfo{journal}{Phys. Rev.} \textbf{\bibinfo{volume}{130}},
  \bibinfo{pages}{2204} (\bibinfo{year}{1963}).

\bibitem[{\citenamefont{Morgan}(1966)}]{morgan66}
\bibinfo{author}{\bibfnamefont{T.~N.} \bibnamefont{Morgan}},
  \bibinfo{journal}{Phys. Rev.} \textbf{\bibinfo{volume}{148}},
  \bibinfo{pages}{890} (\bibinfo{year}{1966}).

\bibitem[{\citenamefont{Yacoby}(1968)}]{yacoby68}
\bibinfo{author}{\bibfnamefont{Y.}~\bibnamefont{Yacoby}},
  \bibinfo{journal}{Phys. Rev.} \textbf{\bibinfo{volume}{169}},
  \bibinfo{pages}{610} (\bibinfo{year}{1968}).

\bibitem[{\citenamefont{Aspnes}(1967)}]{aspnes67}
\bibinfo{author}{\bibfnamefont{D.~E.} \bibnamefont{Aspnes}},
  \bibinfo{journal}{Phys. Rev.} \textbf{\bibinfo{volume}{153}},
  \bibinfo{pages}{972} (\bibinfo{year}{1967}).

\bibitem[{\citenamefont{Enderlein and Keiper}(1967)}]{enderlein67}
\bibinfo{author}{\bibfnamefont{R.}~\bibnamefont{Enderlein}} \bibnamefont{and}
  \bibinfo{author}{\bibfnamefont{R.}~\bibnamefont{Keiper}},
  \bibinfo{journal}{Phys. Stat. Sol.} \textbf{\bibinfo{volume}{19}},
  \bibinfo{pages}{673} (\bibinfo{year}{1967}).

\bibitem[{\citenamefont{French}(1968)}]{french68}
\bibinfo{author}{\bibfnamefont{B.~T.} \bibnamefont{French}},
  \bibinfo{journal}{Phys. Rev.} \textbf{\bibinfo{volume}{174}},
  \bibinfo{pages}{991} (\bibinfo{year}{1968}).

\bibitem[{\citenamefont{Franz}(1958)}]{franz58}
\bibinfo{author}{\bibfnamefont{W.}~\bibnamefont{Franz}},
  \textbf{\bibinfo{volume}{13a}}, \bibinfo{pages}{484} (\bibinfo{year}{1958}).

\bibitem[{\citenamefont{Keldysh}(1965)}]{keldysh65}
\bibinfo{author}{\bibfnamefont{L.}~\bibnamefont{Keldysh}},
  \bibinfo{journal}{JETP} \textbf{\bibinfo{volume}{20}}, \bibinfo{pages}{1307}
  (\bibinfo{year}{1965}).

\bibitem[{\citenamefont{Aspnes and Rowe}(1970)}]{aspnes70}
\bibinfo{author}{\bibfnamefont{D.~E.} \bibnamefont{Aspnes}} \bibnamefont{and}
  \bibinfo{author}{\bibfnamefont{J.~E.} \bibnamefont{Rowe}},
  \bibinfo{journal}{Solid State Commun.} \textbf{\bibinfo{volume}{8}},
  \bibinfo{pages}{1145} (\bibinfo{year}{1970}).

\bibitem[{\citenamefont{Aspnes}(1980)}]{aspnes80}
\bibinfo{author}{\bibfnamefont{D.~E.} \bibnamefont{Aspnes}},
  \emph{\bibinfo{title}{Handbook of semiconductors}}
  (\bibinfo{publisher}{North-Holland}, \bibinfo{address}{Amsterdam},
  \bibinfo{year}{1980}), vol.~\bibinfo{volume}{2}, p. \bibinfo{pages}{109}.

\bibitem[{\citenamefont{Yu and Cardona}(1996)}]{cardona96}
\bibinfo{author}{\bibfnamefont{P.~Y.} \bibnamefont{Yu}} \bibnamefont{and}
  \bibinfo{author}{\bibfnamefont{M.}~\bibnamefont{Cardona}},
  \emph{\bibinfo{title}{Fundamentals of semiconductors}}
  (\bibinfo{publisher}{Springer}, \bibinfo{address}{New York},
  \bibinfo{year}{1996}), chap.~\bibinfo{chapter}{6}, p. \bibinfo{pages}{305}.

\bibitem[{\citenamefont{Seraphin and Hess}(1965)}]{seraphin65}
\bibinfo{author}{\bibfnamefont{B.~O.} \bibnamefont{Seraphin}} \bibnamefont{and}
  \bibinfo{author}{\bibfnamefont{R.~B.} \bibnamefont{Hess}},
  \bibinfo{journal}{Phys. Rev. Lett.} \textbf{\bibinfo{volume}{14}},
  \bibinfo{pages}{138} (\bibinfo{year}{1965}).

\bibitem[{\citenamefont{Garcia}(2006)}]{garcia06d}
\bibinfo{author}{\bibfnamefont{H.}~\bibnamefont{Garcia}},
  \bibinfo{journal}{Phys. Rev. B} \textbf{\bibinfo{volume}{74}},
  \bibinfo{eid}{035212} (\bibinfo{year}{2006}).

\bibitem[{\citenamefont{Garcia and Kalyanaraman}(2006)}]{garcia06a}
\bibinfo{author}{\bibfnamefont{H.}~\bibnamefont{Garcia}} \bibnamefont{and}
  \bibinfo{author}{\bibfnamefont{R.}~\bibnamefont{Kalyanaraman}},
  \bibinfo{journal}{J. Phys. B: At. Mol. Opt. Phys.}
  \textbf{\bibinfo{volume}{39}}, \bibinfo{pages}{2737} (\bibinfo{year}{2006}).

\bibitem[{\citenamefont{Abramowitz and Stegun}(1972)}]{abramowitz}
\bibinfo{editor}{\bibfnamefont{M.}~\bibnamefont{Abramowitz}} \bibnamefont{and}
  \bibinfo{editor}{\bibfnamefont{I.~A.} \bibnamefont{Stegun}}, eds.,
  \emph{\bibinfo{title}{Handbook of {M}athematical {F}unctions}}
  (\bibinfo{publisher}{National Bureau of Standards},
  \bibinfo{address}{Washington, D.C.}, \bibinfo{year}{1972}).

\bibitem[{\citenamefont{Aspnes et~al.}(1968)\citenamefont{Aspnes, Handler, and
  Blossey}}]{aspnes68}
\bibinfo{author}{\bibfnamefont{D.~E.} \bibnamefont{Aspnes}},
  \bibinfo{author}{\bibfnamefont{P.}~\bibnamefont{Handler}}, \bibnamefont{and}
  \bibinfo{author}{\bibfnamefont{D.~F.} \bibnamefont{Blossey}},
  \bibinfo{journal}{Phys. Rev.} \textbf{\bibinfo{volume}{166}},
  \bibinfo{pages}{921} (\bibinfo{year}{1968}).

\bibitem[{\citenamefont{Sheik-Bahae et~al.}(1991)\citenamefont{Sheik-Bahae,
  Hutchings, Hagan, and Van~Stryland}}]{sheikbahae91}
\bibinfo{author}{\bibfnamefont{M.}~\bibnamefont{Sheik-Bahae}},
  \bibinfo{author}{\bibfnamefont{D.}~\bibnamefont{Hutchings}},
  \bibinfo{author}{\bibfnamefont{D.}~\bibnamefont{Hagan}}, \bibnamefont{and}
  \bibinfo{author}{\bibfnamefont{E.}~\bibnamefont{Van~Stryland}},
  \bibinfo{journal}{IEEE J. Quant. Elec.} \textbf{\bibinfo{volume}{27}},
  \bibinfo{pages}{1296} (\bibinfo{year}{1991}).

\bibitem[{\citenamefont{Balu et~al.}(2005)\citenamefont{Balu, Hales, Hagan, and
  Van~Stryland}}]{balu05}
\bibinfo{author}{\bibfnamefont{M.}~\bibnamefont{Balu}},
  \bibinfo{author}{\bibfnamefont{J.}~\bibnamefont{Hales}},
  \bibinfo{author}{\bibfnamefont{D.}~\bibnamefont{Hagan}}, \bibnamefont{and}
  \bibinfo{author}{\bibfnamefont{E.}~\bibnamefont{Van~Stryland}},
  \bibinfo{journal}{OPN} \textbf{\bibinfo{volume}{16}}, \bibinfo{pages}{28}
  (\bibinfo{year}{2005}).

\bibitem[{\citenamefont{He et~al.}(2002)\citenamefont{He, Lin, Prasad, Kannan,
  Vaia, and Tan}}]{he02}
\bibinfo{author}{\bibfnamefont{G.}~\bibnamefont{He}},
  \bibinfo{author}{\bibfnamefont{T.-C.} \bibnamefont{Lin}},
  \bibinfo{author}{\bibfnamefont{P.}~\bibnamefont{Prasad}},
  \bibinfo{author}{\bibfnamefont{R.}~\bibnamefont{Kannan}},
  \bibinfo{author}{\bibfnamefont{R.}~\bibnamefont{Vaia}}, \bibnamefont{and}
  \bibinfo{author}{\bibfnamefont{L.-S.} \bibnamefont{Tan}},
  \bibinfo{journal}{Optics Exp.} \textbf{\bibinfo{volume}{10}},
  \bibinfo{pages}{566} (\bibinfo{year}{2002}).

\bibitem[{\citenamefont{Palik}(1985)}]{palik85}
\bibinfo{author}{\bibfnamefont{E.}~\bibnamefont{Palik}},
  \emph{\bibinfo{title}{Handbook of {O}ptical {C}onstants of {S}olids}}
  (\bibinfo{publisher}{Academic Press}, \bibinfo{address}{NY},
  \bibinfo{year}{1985}).

\bibitem[{\citenamefont{Van~Stryland et~al.}(1985)\citenamefont{Van~Stryland,
  Woodall, Vanherzeele, and Soileau}}]{stryland85}
\bibinfo{author}{\bibfnamefont{E.~W.} \bibnamefont{Van~Stryland}},
  \bibinfo{author}{\bibfnamefont{M.~A.} \bibnamefont{Woodall}},
  \bibinfo{author}{\bibfnamefont{H.}~\bibnamefont{Vanherzeele}},
  \bibnamefont{and} \bibinfo{author}{\bibfnamefont{M.~J.}
  \bibnamefont{Soileau}}, \bibinfo{journal}{Optics Lett.}
  \textbf{\bibinfo{volume}{10}}, \bibinfo{pages}{490} (\bibinfo{year}{1985}).

\bibitem[{\citenamefont{Hurlbut et~al.}(2007)\citenamefont{Hurlbut, Lee, and
  Vodopyanov}}]{hurlbut07}
\bibinfo{author}{\bibfnamefont{W.~C.} \bibnamefont{Hurlbut}},
  \bibinfo{author}{\bibfnamefont{Y.-S.} \bibnamefont{Lee}}, \bibnamefont{and}
  \bibinfo{author}{\bibfnamefont{K.~L.} \bibnamefont{Vodopyanov}},
  \bibinfo{journal}{Optics Lett.} \textbf{\bibinfo{volume}{32}},
  \bibinfo{pages}{668} (\bibinfo{year}{2007}).

\bibitem[{Sop()}]{SopraDataBase}
\emph{\bibinfo{title}{S{OPRA} database}},
  \bibinfo{note}{http://www.sopra-sa.com/}.

\end{thebibliography}

\pagebreak

\section*{Figure captions}

\begin{enumerate}
\item Fig. 1. Comparison of experimental (symbols) and theoretical data
(line) for GaAs. (a) linear absorption \cite{palik85} and fit from
our theory. (b) Nonlinear absorption. (c) Nonlinear refractive index.
The experimental data for $\beta^{(2)}$ and $n_{2}$ was taken from
\cite{stryland85,hurlbut07}. \label{fig:GaAs-data}
\item Fig. 2. Comparison of experimental (symbols) and theoretical data
(line) for ZnSe. (a) linear absorption \cite{SopraDataBase} and fit
from our theory. (b) Nonlinear absorption. (c) Nonlinear refractive
index. The experimental data for $\beta^{(2)}$ and $n_{2}$ was taken
from \cite{balu05}. \label{fig:ZnSe-data}
\end{enumerate}
\pagebreak

\section*{Table captions}

\begin{enumerate}
\item Table of quantities used in the calculations. $m_{c}$, $m_{v}$,
$m_{o}$ and $n_{o}$ are the known conduction electron mass, hole
mass, electron mass and refractive index. $E_{g}$ and $\alpha_{b}^{'}$
are energy gap and fitting parameter extracted from the linear absorption
spectrum. \label{tab:Table-of-quantities}
\pagebreak
\end{enumerate}

\begin{flushright}%
\begin{figure}[t]
\subfigure[]{\includegraphics[height=2.5in,keepaspectratio]{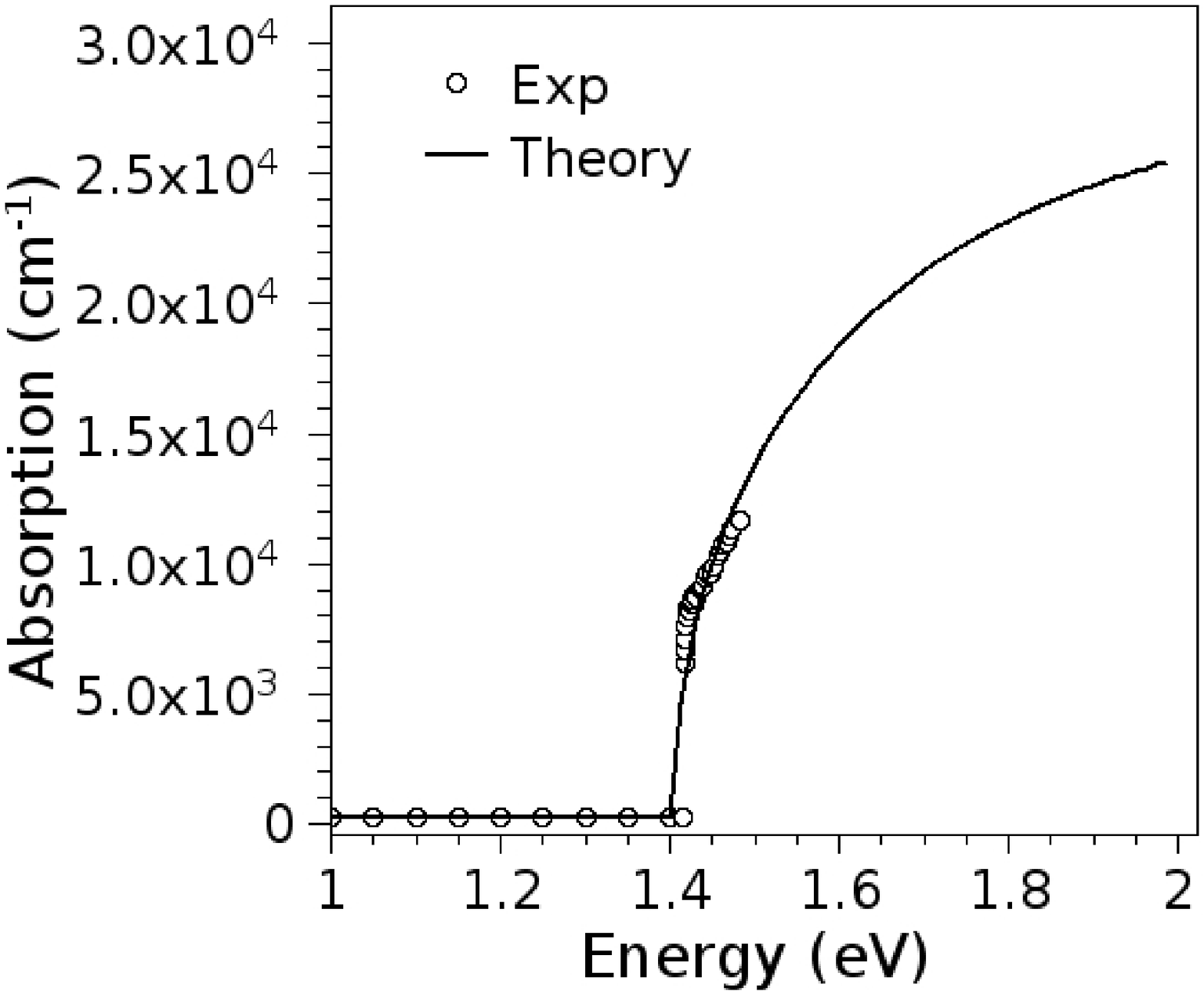}}

\subfigure[]{\includegraphics[height=2.5in,keepaspectratio]{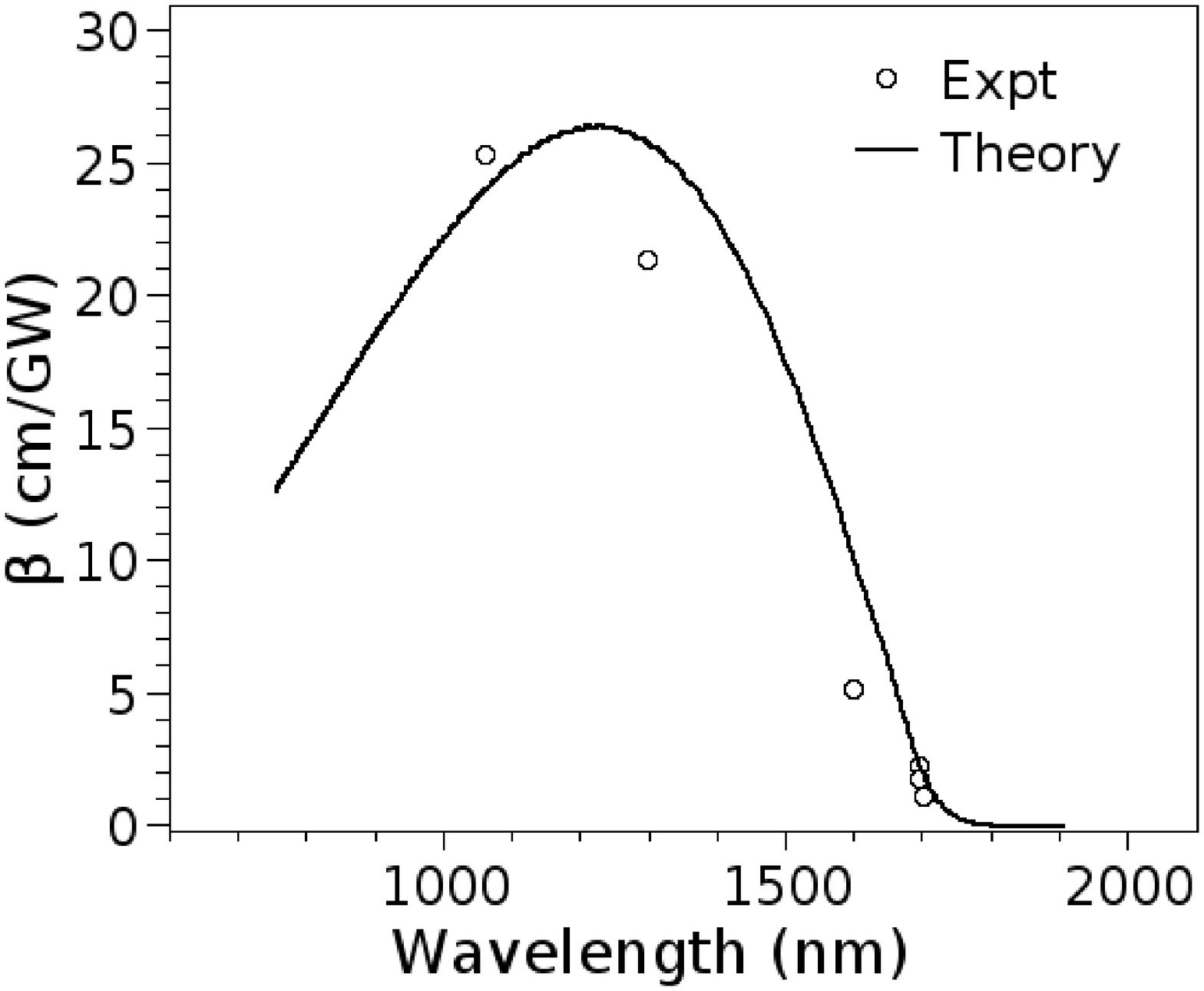}}

\subfigure[]{\includegraphics[height=2.5in,keepaspectratio]{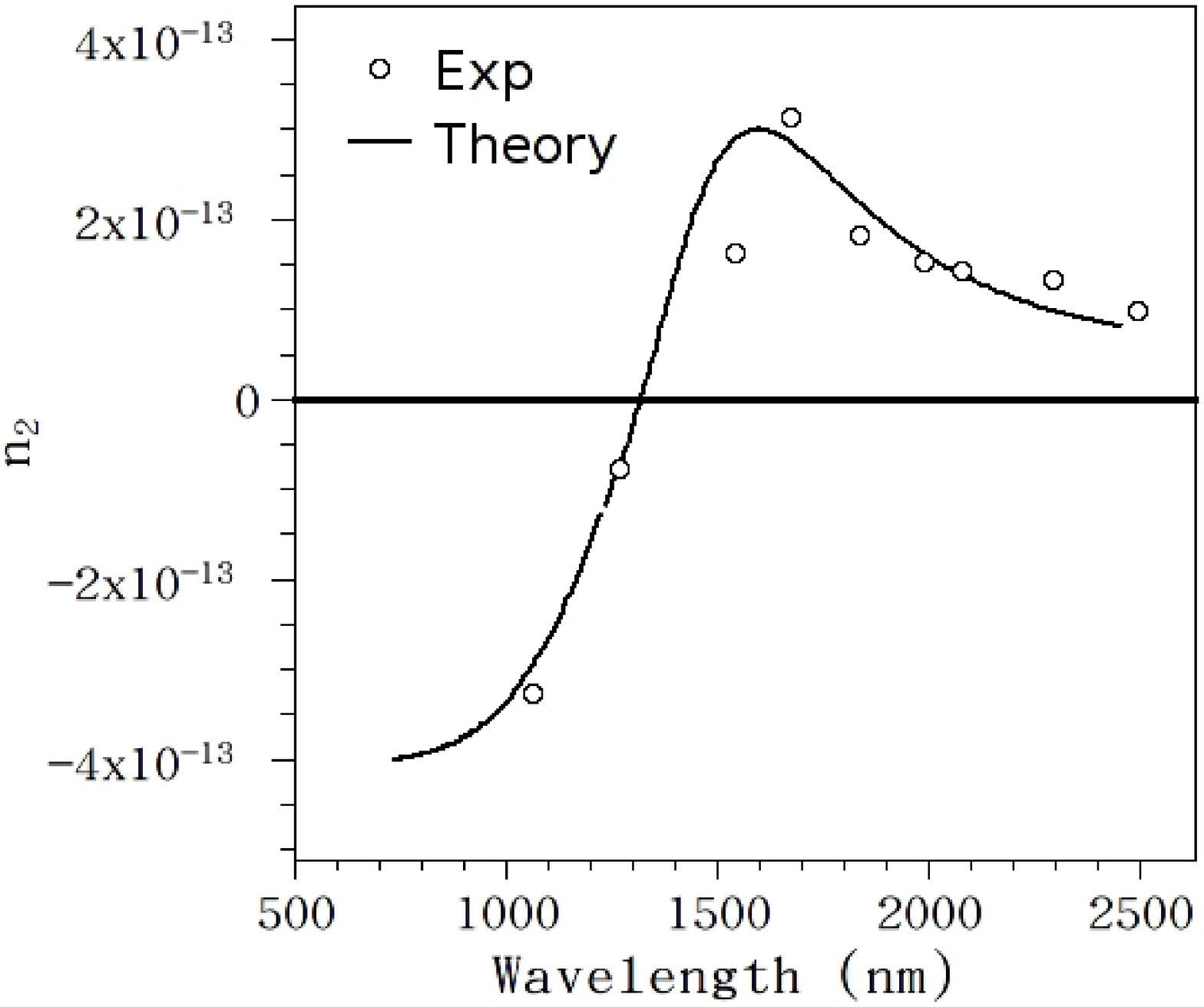}}

\caption{Comparison of experimental (symbols) and theoretical data (line)
for GaAs. (a) linear absorption \cite{palik85} and fit from our theory.
(b) Nonlinear absorption. (c) Nonlinear refractive index. The experimental
data for $\beta^{(2)}$ and $n_{2}$ was taken from \cite{stryland85,hurlbut07}.
\label{fig:GaAs-data}}
\end{figure}
\par\end{flushright}

\pagebreak

\begin{center}%
\begin{figure}[t]
\subfigure[]{\includegraphics[height=2.5in,keepaspectratio]{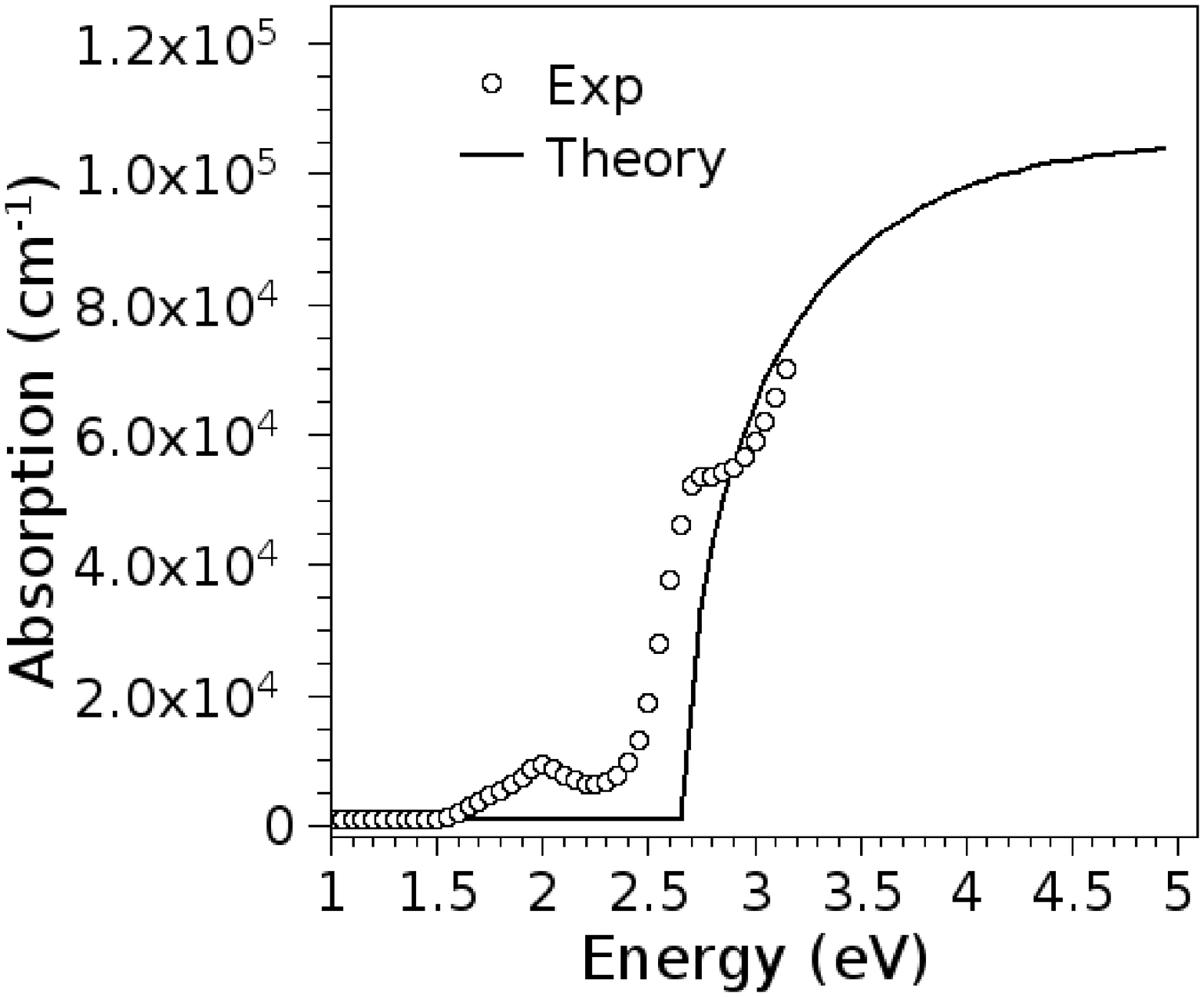}}

\subfigure[]{\includegraphics[height=2.5in,keepaspectratio]{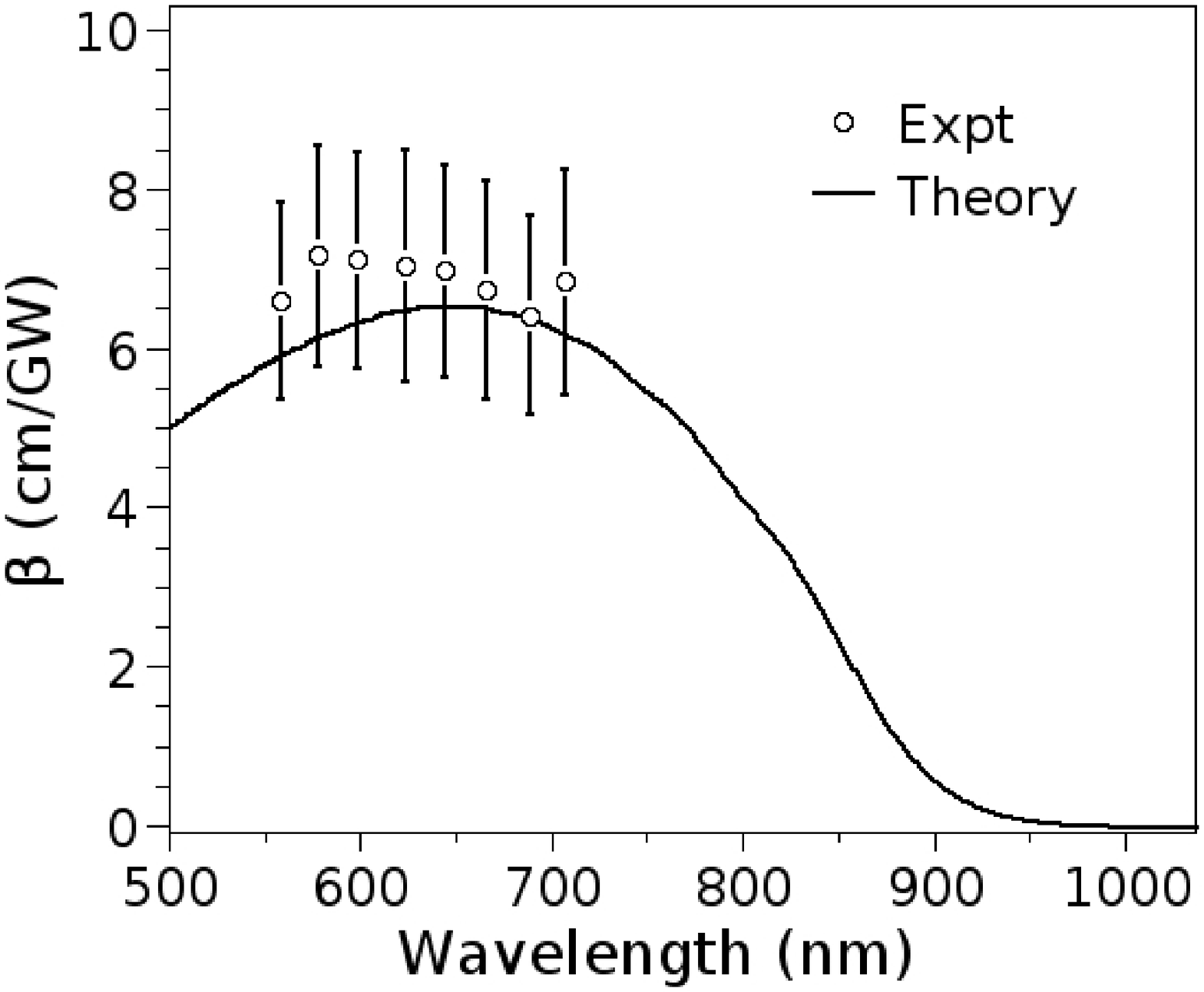}}

\subfigure[]{\includegraphics[height=2.5in,keepaspectratio]{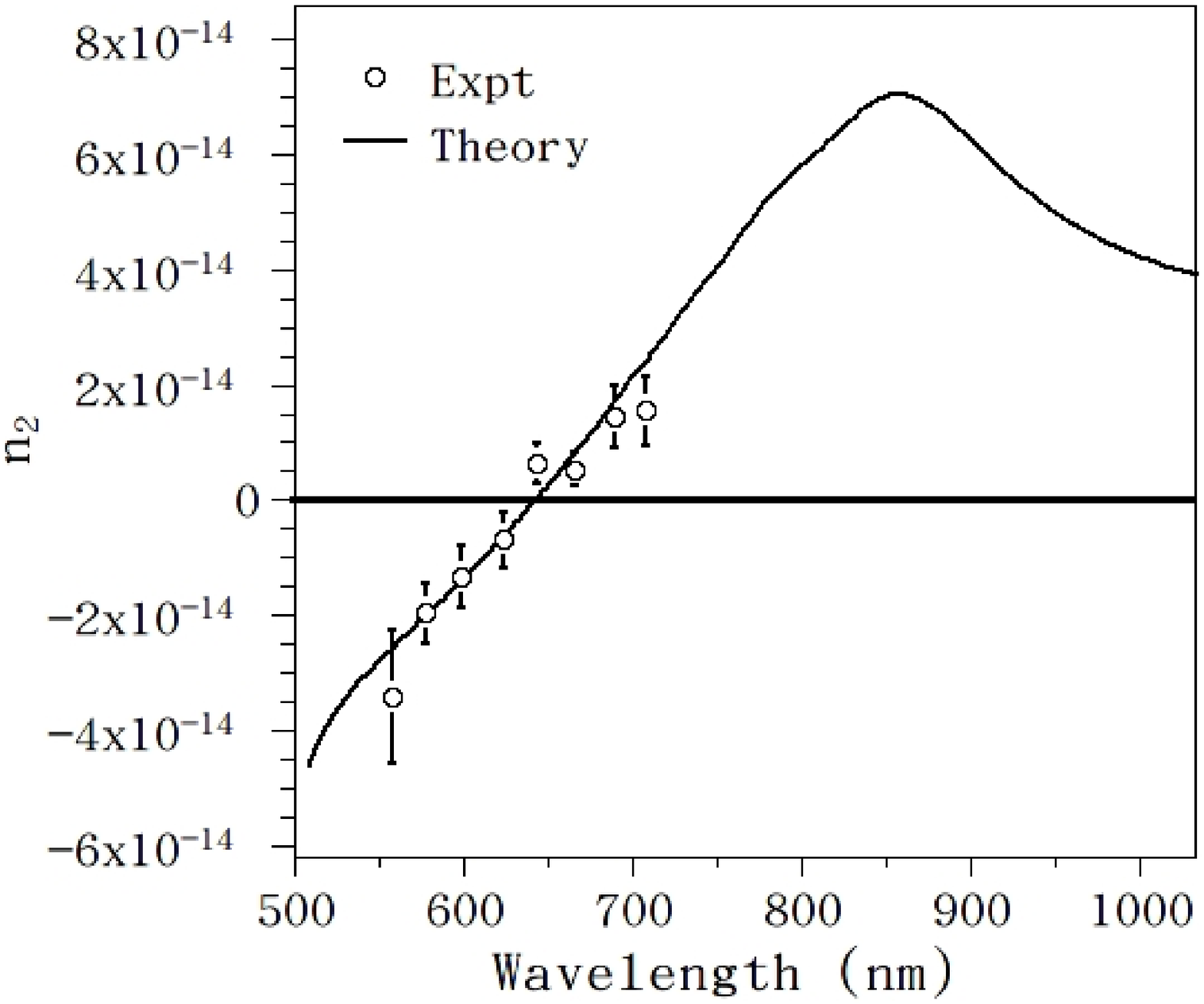}}

\caption{Comparison of experimental (symbols) and theoretical data (line)
for ZnSe. (a) linear absorption \cite{SopraDataBase} and fit from
our theory. (b) Nonlinear absorption. (c) Nonlinear refractive index.
The experimental data for $\beta^{(2)}$ and $n_{2}$ was taken from
\cite{balu05}. \label{fig:ZnSe-data}}
\end{figure}
\pagebreak\par\end{center}

\begin{center}%
\begin{table}[tbh]
\begin{tabular}{|c|c|c|}
\hline 
&
GaAs&
ZnSe\tabularnewline
\hline
\hline 
$E_{g}$&
1.403 eV&
$2.67\, eV$\tabularnewline
\hline 
$m_{c}$&
$0.067m_{o}$&
$0.17m_{o}$\tabularnewline
\hline 
$m_{v}$&
$0.68m_{o}$&
$1.44m_{o}$\tabularnewline
\hline 
$n_{o}$&
3.42&
2.48\tabularnewline
\hline 
$\alpha_{b}^{'}$&
$2.43\times10^{-6}(\frac{eV}{gm})^{1/2}erg$&
$9.51\times10^{-12}(\frac{eV}{gm})^{1/2}erg$\tabularnewline
\hline
\end{tabular}

\caption{Table of quantities used in the calculations. $m_{c}$, $m_{v}$,
$m_{o}$ and $n_{o}$ are the known conduction electron mass, hole
mass, electron mass and refractive index. $E_{g}$ and $\alpha_{b}^{'}$
are energy gap and fitting parameter extracted from the linear absorption
spectrum. \label{tab:Table-of-quantities}}
\end{table}
\par\end{center}
\end{document}